\documentstyle[psfig]{mn2e}
\newcommand{\etal}{{et al.~}}

\newcommand{\kms}{\>{\rm km}\,{\rm s}^{-1}}

\newcommand{\mpc}{\>{\rm Mpc}}

\newcommand{\beq}{\begin{equation}}
\newcommand{\eeq}{\end{equation}}

\newcommand{\mpci}{\ $h$\ Mpc$^{-1}$\ }

\newcommand{\apj}{ApJ}

\newcommand{\aj}{AJ}
\newcommand{\mnras}{MNRAS}

\newcommand{\nat}{Nature}

\newdimen\hssize
\newdimen\hdsize 
\def\beq{\begin{equation}}
\def\eeq{\end{equation}}
\def\bey{\begin{eqnarray}}
\def\eey{\end{eqnarray}}

\def\mpc{\,h^{-1}\ {\rm {Mpc}}}
\def\mpci{\,h\ {\rm {Mpc}}^{-1}}

\def\kms{\,{\rm {km\, s^{-1}}}}

\def\gs{\mathrel{\raise1.16pt\hbox{$>$}\kern-7.0pt
\lower3.06pt\hbox{{$\scriptstyle \sim$}}}}
\def\ls{\mathrel{\raise1.16pt\hbox{$<$}\kern-7.0pt
\lower3.06pt\hbox{{$\scriptstyle \sim$}}}}
\def\gtsima{$\; \buildrel > \over \sim \;$}
\def\ltsima{$\; \buildrel < \over \sim \;$}
\def\prosima{$\; \buildrel \propto \over \sim \;$}
\def\gsim{\lower.5ex\hbox{\gtsima}}
\def\lsim{\lower.5ex\hbox{\ltsima}}
\def\simgt{\lower.5ex\hbox{\gtsima}}
\def\simlt{\lower.5ex\hbox{\ltsima}}
\def\simpr{\lower.5ex\hbox{\prosima}}

\begin{document}

\title
[The dependence of PVD on galaxy properties]
{The dependence of the pairwise velocity dispersion on galaxy properties}
\author[C. Li \etal]
{Cheng Li${^{1,2,3}}$ \thanks{E-mail: leech@ustc.edu.cn},
 Y.P. Jing${^{1}}$, Guinevere Kauffmann${^{3}}$, Gerhard B\"orner$^{3}$,
\newauthor Simon D.M. White$^{3}$, F.Z. Cheng$^{2}$\\
${^1}$The Partner Group of MPI f\"ur Astrophysik,
      Shanghai Astronomical Observatory,
      Nandan Road 80, Shanghai 200030, China \\
${^2}$Center for Astrophysics, University of Science
      and Technology of China, Hefei, Anhui 230026, China \\
${^3}$Max-Planck-Institut f\"ur Astrophysik,
      Karl-Schwarzschild-Strasse 1, 85748 Garching, Germany}
                                                                                
\date{
Accepted ........
Received .......;
in original form ......}
\pubyear{2006}
\maketitle

\begin{abstract}
We present measurements of the pairwise velocity dispersion (PVD) for
different classes of galaxies in the Sloan Digital Sky Survey (SDSS). 
For a sample of about 200,000 galaxies with redshifts in the interval
$0.01 < z < 0.3 $, and r-band magnitudes $M_{^{0.1}r}$ between $-16$
and $-23$, we study the dependence of the PVD on galaxy properties
such as luminosity, stellar mass  ($M_\ast$), colour ($g-r$), 4000\AA\
break strength (D$_{4000}$), concentration index ($C$), and stellar
surface mass density ($\mu_\ast$). The luminosity dependence of the
PVD is in good agreement with the results of Jing \& B\"orner.
for the 2dFGRS catalog, once the photometric bandpass difference
between the two surveys is taken into account. The value of $\sigma_{12}$
measured at $k=1 \mpci$ {\em decreases} as a function of increasing
galaxy luminosity for galaxies fainter than $L^\ast$, before increasing
again for the most luminous galaxies in our sample. 
Each of the galaxy subsamples selected according to luminosity or
stellar mass is divided into two  further subsamples according to colour,
D$_{4000}$, $C$ and $\mu_\ast$. We find that galaxies with redder colours
and higher D$_{4000}$, $C$, and $\mu_\ast$ values have larger PVDs on all
scales and at all luminosities/stellar masses.  The dependence of the
PVD on parameters related to recent star formation (colour, D$_{4000}$)
is stronger than on parameters related to galaxy structure ($C$, $\mu_\ast$), 
especially on small scales and for faint galaxies. The reddest galaxies
and galaxies with high surface mass densities and intermediate
concentrations have the highest pairwise peculiar velocities, i.e.
these move in the strongest gravitational fields.  We conclude that
the faint red population located in rich clusters is responsible for the
high PVD values that are measured for low-luminosity galaxies on small
scales.
\end{abstract}


\section {Introduction}
The pairwise peculiar velocity dispersion (PVD) measures the relative
motions of galaxies, and so reflects the action of the local
gravitational fields. The PVD is thus an important quantity for
probing the mean mass density $\Omega_0$ of the Universe and the
clustering power on small scales. It is widely used to constrain
cosmogonic models and galaxy formation recipes (Davis \etal 1985).
As a well-defined statistical quantity, the PVD can be estimated from
redshift surveys, either by modeling redshift distortions in the
two-point correlation function (2PCF), or by measuring the
redshift-space power spectrum (Davis \& Peebles 1983; Jing \& B\"orner 2001b).

The first method relies on the fact that the peculiar motions of
galaxies affect only their radial distances in redshift space.  Thus
the information for peculiar velocities along the line of sight can be
recovered by modeling the redshift-space 2PCF 
$\xi^{(s)}(r_p,\pi)$ as
a convolution of the real-space 2PCF $\xi(r)$ with the distribution
function of the pairwise velocity $f(v_{12})$:
\begin{equation}\label{eqn:xipv}
\xi^{(s)}(r_p,\pi)=
\int f(v_{12})\xi\left(\sqrt{r_p^2+(\pi-v_{12})^2}\right)dv_{12},
\end{equation}
where $v_{12}=v_{12}(r_p,\pi)$ is the pairwise peculiar velocity;
$r_p$ and $\pi$ are the separations perpendicular and parallel
to the line of sight.  The real-space correlation function $\xi(r)$ is
usually inferred from the projected 2PCF $w_p(r_p)$, which is a simple
Abel transform of $\xi(r)$. However, the form of $f(v_{12})$ is not
known from a rigorous theory. Based on observational
(Davis \& Peebles 1983; Fisher \etal 1994)
and theoretical considerations (e.g. Diaferio \& Geller 1996;
Sheth 1996), an exponential form is usually adopted:
\begin{equation}\label{eqn:expfv}
f(v_{12})=\frac{1}{\sqrt{2}\sigma_{12}}
\exp\left(-\frac{\sqrt{2}}{\sigma_{12}}
\left|v_{12}-\overline{v_{12}}\right|\right)
\end{equation}
where $\overline{v_{12}}$ is the mean and $\sigma_{12}$ is the
dispersion of the one-dimensional peculiar velocicites.
Assuming an infall model for $\overline{v_{12}}(r)$, the pairwise velocity dispersion
$\sigma_{12}$ can then be estimated as a function of projected
separation $r_p$ by comparing the observed $\xi^{(s)}(r_p,\pi)$ with
the modeled one. The form of the infall model usually adopted is based on
the self-similar solution and is a good approximation to the real
infall pattern in CDM models with $\beta\equiv
\sigma_8\Omega_0^{0.6}\approx0.5$ (Jing, Mo \& B\"orner 1998, hereafter JMB98).
Using this method and various redshift samples of galaxies, the
measurement of PVD has been carried out  by many authors
(Davis \& Peebles 1983; Mo, Jing \& B\"orner 1993; Fisher \etal 1994;
Zurek \etal 1994; Marzke \etal 1995; Somerville, Davis \& Primack 1997).
There have been significant variations in the results of these studies. 
In many cases the measurement errors were underestimated and there were
also problems due to ``cosmic variance'' as a result of the limited
volume contained in the early redshift surveys (Mo, Jing \& B\"orner 1997).
As a result, the early determinations are of limited power in
constraining large-scale stucture  theories.  The first accurate determination of the PVD
($\sigma_{12}=570\pm80\kms$ at a projected separation $r_p=1\mpc$)
was presented by JMB98 using the Las Campanas redshift Survey (LCRS),
and has been verified by Zehavi \etal (2002) with the early data release
of the Sloan Digital Sky Survey (SDSS) and by Hawkins \etal (2003)
with the two-degree field galaxy redshift survey (2dFGRS). This value
is substantially higher than the earlier results based on smaller
surveys ($340\pm40\kms$) given by Davis \& Peebles (1983).
The large difference in these results is due to the fact that the value of
the PVD is very
sensitive to the presence (or absence) of rich clusters in a sample
(Mo, Jing \& B\"orner 1993), and so a sample as large as LCRS is needed
in order to give reliable estimates.

The PVD of galaxies can also be determined from the redshift-space
power spectrum. Although the power spectrum is just the Fourier
transform of the 2PCF, it is sometimes advantageous to work with the
power spectrum instead of with the 2PCF.  The advantage of using the
redshift-space power spectrum to determine the PVD is that it is simple and
accurate to model the infall effect (Jing \& B\"orner 2004, hereafter JB04).
It is also not necessary to assume a functional form for the real
space $P(k)$, as is usually required for the 2PCF. The relation
between the power spectrum in redshift space $P^{(s)}(k,\mu)$ and that
in real space $P(k)$ can be written as
(Peacock \& Dodds 1994; Cole, Fisher \& Weinberg 1995):
\begin{equation}\label{eqn:redpow}
P^{(s)}(k,\mu) = P(k)(1+\beta \mu^2)^2 D(k \mu \sigma_{12}(k)).
\end{equation}
Here $k$ is the wavenumber, $\mu$ the cosine of the angle between the
wavevector and the line of sight, and $\beta$ the linear redshift
distortion parameter.  The first multiplies after the power spectrum
on the right hand side of Eq.(\ref{eqn:redpow}) is the Kaiser linear
compression effect (Kaiser 1987), and the term $D$ is the damping
effect caused by the random motion of the galaxies.  The
damping function $D$, which should generally depend on $k$, $\mu$ and
$\sigma_{12}(k)$, was found by Jing \& B\"orner (2001a) to be a
function of one variable $k\mu\sigma_{12}(k)$ only. They also found
that, although the functional form of $D[k\mu\sigma_{12}(k)]$ depends
on the cosmological model and bias recipes, for small $k$ (large
scales) where $D>0.1$, $D[k\mu\sigma_{12}(k)]$ can be approximately
expressed by the Lorentz form,
\begin{equation}\label{eqn:damping}
D(k \mu \sigma_{12}(k)) = \frac{1}{1+ \frac{1}{2}k^2 \mu^2 \sigma_{12}(k)^2}\,.
\end{equation}
We will use equations (\ref{eqn:redpow}) and (\ref{eqn:damping}) to
derive $\sigma_{12}$ and $P(k)$ from a measurement of the
redshift-space power spectrum $P^{(s)}(k,\mu)$.  However, when
applying this method, there are a couple of important points that
should be kept in mind.  First, the Lorentz form is the Fourier
transform of the exponential form of $f(v_{12})$ when $\sigma_{12}$ is
a constant and $\overline{v_{12}}$ is zero. This means that
$\sigma_{12}(k)$ drived in the Fourier space is generally not the same
as $\sigma_{12}(r)$ defined in the configuration space. Furthermore,
even if $\sigma_{12}$ is a constant, the redshift power spectrum of
equation (\ref{eqn:redpow}) may lead to an unphysical distribution of
pairwise velocities in the configuration space when the infall is not
zero (i.e. $\beta\neq 0$) (Scoccimarro 2004). Nevertheless, based on
extensive tests of numerical simulations, Jing \& B\"orner (2001a)
demonstrated that, $\sigma_{12}(r)$ and $\sigma_{12}(k)$ have similar
dependences on scale, and are different only by 15 per cent if $r=1/k$
is used in the comparison.  Therefore, the quantity $\sigma_{12}(k)$
can still be regarded as a good indicator for the pairwise velocity
dispersion.  Second, it is usually difficult to determine
$\sigma_{12}$ and $\beta$ simultaneously, because there exists a
strong degeneracy in determining these parameters (Peacock \etal 2001)
from $P^{(s)}(k,\mu)$ at small scales.  Moreover, there could be some
luminosity dependence in $\beta$.  JB04 investigated this by using the
luminosity dependence of the bias parameter $b$ given in (Norberg
\etal 2002), and showed that their determinations of the PVD are
robust to reasonable changes of the $\beta$ values.  Thus a viable way
to determine the PVD from $P^{(s)}(k,\mu)$ is to fix $\beta$ at a
reasonable estimate and then determine $P(k)$ and $\sigma_{12}$ from
the data.  In this way, the real-space power spectrum $P(k)$ is mainly
determined from $P^{(s)}(k,\mu=0)$, and the measured PVD $\sigma_{12}$
is usually a function of $k$.  Adopting Eq. (\ref{eqn:damping}) for
the damping function and setting $\beta=0.45$, Jing \& B\"orner
(2001b) measured the PVD $\sigma_{12}(k)$ for the LCRS and found that
the measured PVD is consistent with the result reported by JMB98 from
the redshift space 2PCF measurement.

Both of the methods described above for determining the PVD of
galaxies are based on the measurement of galaxy clustering.
Previous studies of galaxy clustering in the local Universe
have established that it depends on a variety of quantities, 
including luminosity (e.g. Norberg \etal 2001;
Zehavi \etal 2002, 2005), colour (e.g. Zehavi \etal 2002, 2005),
morphology (e.g. Zehavi \etal 2002; Goto \etal 2003),
and spectral type (e.g. Norberg \etal 2002; Budav{\' a}ri \etal 2003;
Madgwick \etal 2003).
It is thus expected that there should be some dependence of the
galaxy PVD on these physical properties.  Using the 2dFGRS and the
method based on the redshift-space power spectrum, JB04 present the
first determination of the PVD for galaxies in different luminosity
intervals.  The analysis leads to a surprising discovery:  the
relative velocities of the faint galaxies are very high, around
700$\kms$, reaching values similar to those found
for the brightest galaxies in the sample. 
The relative velocities at intermediate luminosities $M^\ast-1$
($M^\ast$ is the characteristic luminosity of the Schechter [1976] function)
exhibit a well-defined minimum near 400$\kms$. This discovery
indicates that faint galaxies, and the brightest ones, are both preferentially
in massive haloes of galaxy cluster size, but most
 $M^\ast$ galaxies are in galactic scale haloes. Such a
luminosity dependence of the PVD is not reproduced by the halo
occupation model of Yang \etal (2003), although this model
reproduces well the luminosity dependence of the clustering.

\begin{figure*}
\vspace{-3.2cm}
\centerline{\psfig{figure=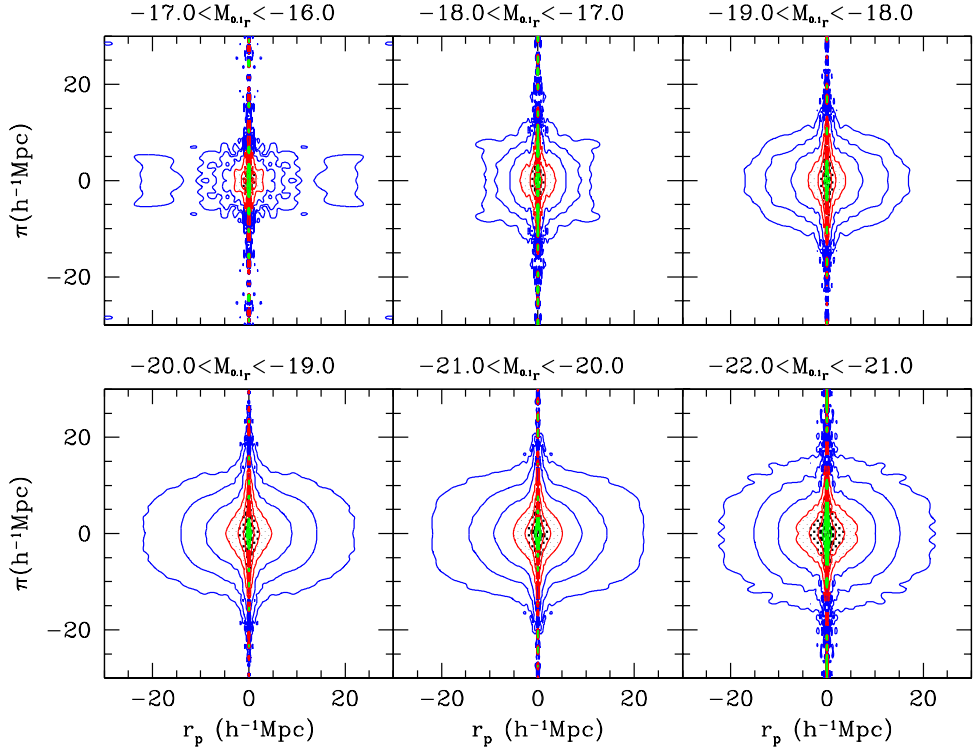,width=14cm}}
\vspace{-0.8cm}
\caption{The redshift space two-point correlation function $\xi^{(s)}(r_p,\pi)$
in different luminosity intervals, as indicated. The contour levels are increased
by factors of 2 from the lowest (0.1875) to the highest (48.0).}
\label{fig:xipv_abm}
\end{figure*}
\begin{figure*}
\vspace{-3.2cm}
\centerline{\psfig{figure=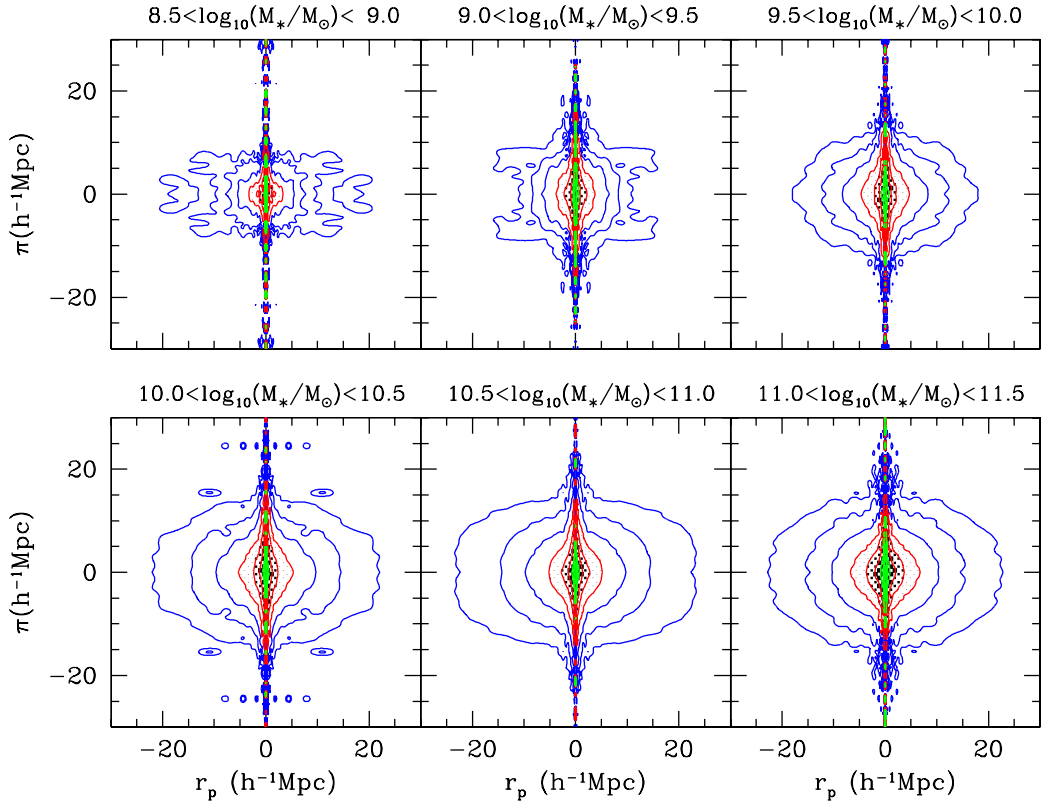,width=15cm}}
\vspace{-0.8cm}
\caption{The redshift space two-point correlation function $\xi^{(s)}(r_p,\pi)$
in different stellar mass intervals, as indicated.  The contour levels are increased
by factors of 2 from the lowest (0.1875) to the highest (48.0).}
\label{fig:xipv_smass}
\end{figure*}

In this paper, we use the second data release of the SDSS to
examine the dependence of the PVD not only on 
luminosity, but also, for the first time, on other physical properties such as the stellar
mass, star formation rate, stellar surface mass density
and concentration. These dependences will
provide useful clues about how the luminosity dependence of the PVD is
produced in the real Universe. As we will see, the high PVD values measured 
at faint luminosities can be explained by a population of red galaxies
located in rich groups and clusters, which is
otherwise sub-dominant in many other clustering measures such as the
2PCF and measurements of galaxy-galaxy lensing. The results are
consistent with the conjecture made by JB04 for this population.
In a separate paper (Li \etal 2006, hereafter Paper I),
we have measured the 2PCF for the galaxy samples studied in
this paper. Taken together, our determinations of the 2PCF and the PVD   
should be extremely useful for constraining models of galaxy formation.

In the following section we briefly describe the measurements
of the 2PCFs of galaxies that are presented in Paper I.
In \S3 we describe the method and the results
for determining the PVD from the redshift power spectrum.
We summarize our results in the final section.

Throughout this paper, we assume a cosmological model with the density
parameter $\Omega_0=0.3$ and the cosmological constant
$\Lambda_0=0.7$. 
To avoid the $-5\log_{10}h$ constant, the Hubble's costant $h=1$,
in units of 100 kms$^{-1}$Mpc$^{-1}$, is assumed throughout this paper when
computing absolute magnitudes.
In this paper, the quantities with a superscript
asterisk are those at the characteristic luminosity/mass
(e.g. characteristic luminosity $L^\ast$), while the quantities with
a subscript asterisk are those of stars (e.g. stellar mass $M_\ast$).

\section{Measurements of the two-point correlation functions}
In Paper I, we  described in detail our procedure for
constructing the observational and random samples
and for 
measuring the two-point correlation functions in redshift space.
The samples analyzed here are exactly the same
as those presented in Paper I and the reader is referred
to this paper for more details.

The redshift-space 2PCF $\xi^{(s)}(r_p,\pi)$ for each subsample
is measured using the Hamilton (1993) estimator,
\begin{equation}
\xi^{(s)}(r_p,\pi) = \frac{4DD(r_p,\pi)RR(r_p,\pi)}{[DR(r_p,\pi)]^2}-1.
\end{equation}
Here $DD(r_p,\pi)$ is the count of data-data pairs with
perpendicular separations in the bins $r_p \pm 0.5\Delta r_p$
and with radial separations in the bins $\pi \pm 0.5\Delta\pi$,
$RR(r_p,\pi)$ and $DR(r_p,\pi)$ are similar counts of random-random
and data-random pairs, respectively.
When computing 2PCFs, some possible biases such as the
variance in mass-to-light ratio and fibre collisions have
been carefully corrected (see Paper I for a detailed description).
Figs.\ref{fig:xipv_abm} and \ref{fig:xipv_smass}
show the contours of $\xi^{(s)}(r_p,\pi)$ for galaxies
of different luminosities and of different stellar masses respectively.
Both the effect of redshift space distortions on small scales
(often called the Finger-of-God effect) and the infall effect
(Kaiser 1987) on large scales are clearly visible:
on small scales $\xi^{(s)}(r_p,\pi)$ is stretched in the $\pi$-direction
and on large scales the contours are squashed along the line-of-sight
direction.

In Paper I, we presented the projected two-point correlation function
$w_p(r_p)$ which is estimated from $\xi^{(s)}(r_p,\pi)$. With this
estimator, we investigated the clustering properties of galaxies
with various physical properties. In the next section, we will present
their real-space power spectra and pairwise velocity dispersions.

\begin{figure*}
\vspace{-7.2cm}
\centerline{\psfig{figure=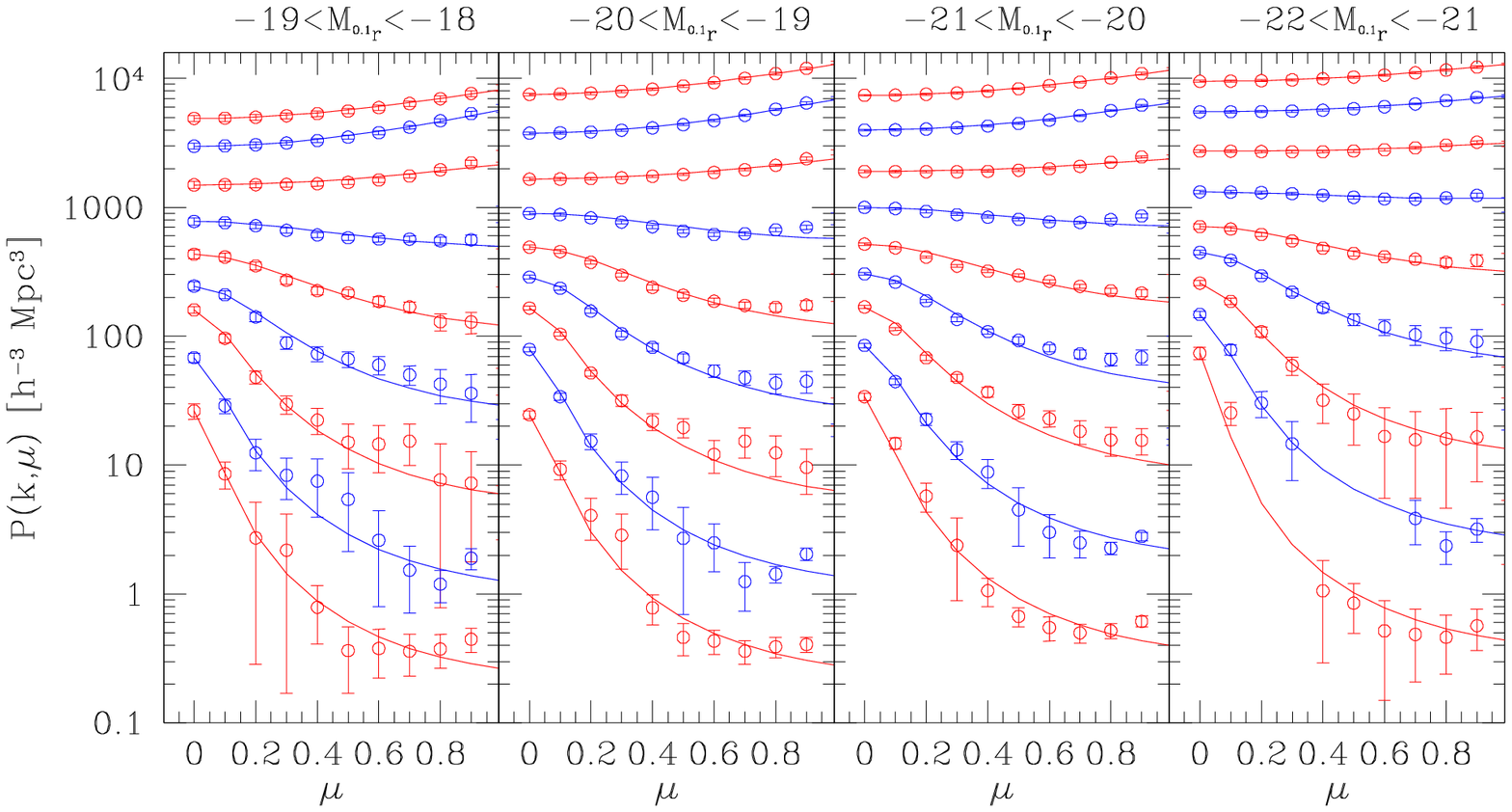,width=\hdsize}}
\vspace{-1.0cm}
\caption{The redshift space power spectrum $P^{(s)}(k,\mu)$ 
({\it symbols with error bars}) for galaxies in different luminosity
intervals, as indicated above each panel. 
The lines are the best fits of Eq.(\ref{eqn:redpow})
to the data. In each panel, from top to bottom, the wavenumber $k$
is 0.1, 0.16, 0.25, 0.4, 0.63, 1.0, 1.58, 2.51, and 3.98 $h$ Mpc$^{-1}$,
respectively.}
\label{fig:powpre}
\end{figure*}
\section{The power spectrum and the pairwise velocity dispersion}

\subsection{Method}
Following the method of JB04, we obtain for each subsample
the redshift-space power spectrum as
\begin{equation}
\begin{array}{rcl}
P^{(s)}(k,\mu) & = & 2\pi \sum_{i,j} \Delta \pi_{i} r_{p,j}^2 \Delta \ln r_{p,j}
\xi^{(s)}(r_{p,j},\pi_i) \\
& & \cos(k_\pi \pi_i) J_0(k_p r_{p,j})W_g(r_{p,j},\pi_i),
\end{array}
\label{eq6}
\end{equation}
where $J_0$ is the zeroth-order Bessel function (Jing \& B\"orner 2001b);
$k_p$ and $k_\pi$ are the wavenumbers perpendicular and parallel to the
line of sight, and are related to $k$ and $\mu$ with
\begin{equation}
k=\sqrt{k_p^2+k_\pi^2},\mbox{\ \ \ }\mu=k_\pi/k.
\end{equation}
Here $\pi_{i}$ runs from $-40$ to $40\mpc$ with $\Delta
\pi_{i}=1\mpc$ and $r_{p,j}$ from $0.1$ to $50\mpc$ with $\Delta \ln
r_{p,j}=0.23$.
$W_g$ is the Gaussian window function, which is used to improve
the measurement by down-weighting $\xi^{(s)}(r_p,\pi)$ at the
larger scales and has the form
\begin{equation}
W_g(r_p,\pi)=\exp\left(-\frac{r_p^2+\pi^2}{2S^2}\right).
\end{equation}
The smoothing scale is set to be $S=20\mpc$ following JB04. With these
parameters, JB04 showed that the real space $P(k)$ and the PVD
$\sigma_{12}(k)$ can be reliably measured at scales $k>0.1\mpci$ and
$k>0.2\mpci$ respectively.

From the measurement of the redshift-space power spectrum
$P^{(s)}(k,\mu)$, we can determine the real space power spectrum
$P(k)$ and the PVD $\sigma_{12}$ simultaneously by modeling the measured
$P^{(s)}(k,\mu)$ using Eq.(\ref{eqn:redpow}).  As discussed in \S1, it
is difficult to simultaneously derive $\beta$, $\sigma_{12}$ and
$P(k)$ from $P^{(s)}(k,\mu)$.  We therefore fix $\beta=0.45$ as a
reasonable estimate (Tegmark \etal 2004) and determine $P(k)$
and $\sigma_{12}$ from the data.  We will show in the next section
that the determinations of the PVD are robust to reasonable changes of
the $\beta$ values.

\subsection{The dependence on luminosity}

\begin{figure*}
\vspace{-4.5cm}
\centerline{\psfig{figure=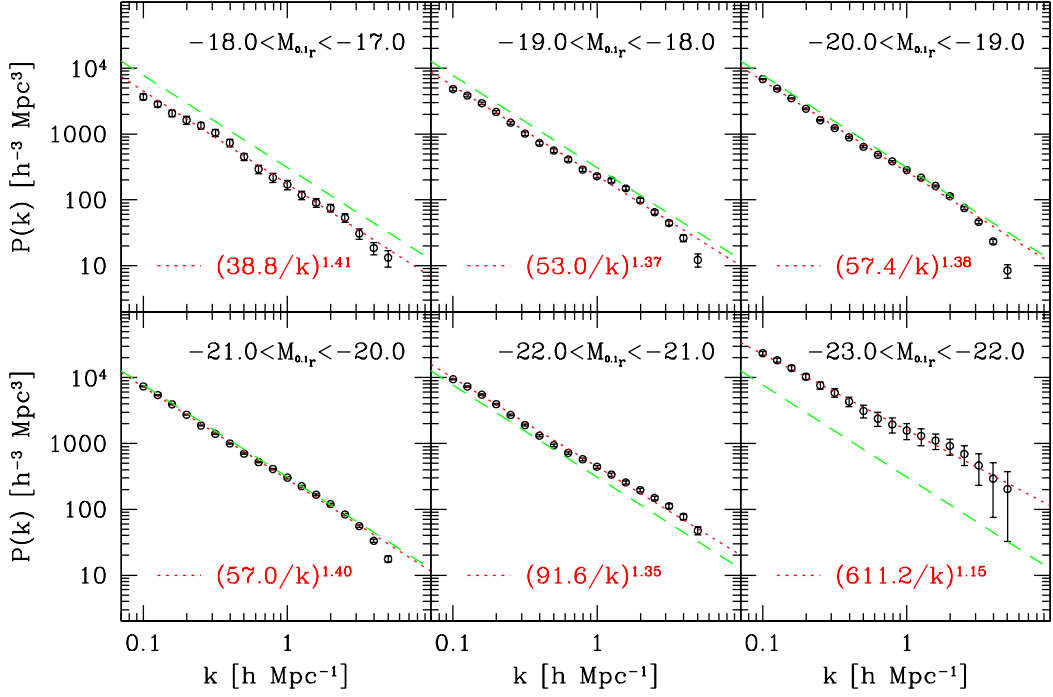,width=15cm}}
\vspace{-0.8cm}
\caption{The real space power spectrum $P(k)$ for galaxies
in different luminosity intervals, as indicated.
The dotted lines are the power-law fits with the best-fitting
parameters indicated in each panel.
To guide the eye, the power spectrum $P(k)=(60/k)^{1.4}$ is
plotted as dashed line in every panel.
}
\label{fig:powreal}
\end{figure*}

In Fig.\ref{fig:powpre} the symbols with error bars show the
measured power spectrum in redshift space $P^{(s)}(k,\mu)$. The
four panels correspond to four different luminosity intervals
(samples L5, L7, L9, and L11 in Table 1 of Paper I).
In each panel, the values of $k$ range from
0.1$\mpci$ at the top, to 4.0$\mpci$ at the bottom, with an increment
of $\Delta\log_{10}k=0.2$.  The error bars plotted in this figure and
in all subsequent figures are estimated by the bootstrap resampling
technique (Barrow, Bhavsar, \& Sonoda 1984).  We generate 100
bootstrap samples and compute the power spectrum for each sample using
the weighting scheme (but not the approximate formula) given by Mo,
Jing, \& B\"orner (1992).  The errors are the scatter of the power
spectra among these bootstrap samples.  The solid lines are the best
fits obtained by applying Eq.(\ref{eqn:redpow}) to the data.  The
real-space power spectrum $P(k)$, determined from the modelling with Eq.(\ref{eqn:redpow}),
is displayed in Fig.\ref{fig:powreal}
for 6 luminosity samples (Samples L5, L7, L9, L10, L11 and L13 of Table 1 in Paper I).
To see the systematic change
with the luminosity, a function $P(k)=(60/k)^{1.4}$ is plotted as the
dashed line in each panel. From both figures, it is seen that there
exists a luminosity dependence of the real-space power spectrum:
$P^{(s)}(k,\mu=0)$ and $P(k)$ increase with increasing luminosity on
all scales, with the luminosity dependence becoming stronger
for bright galaxies and on small scales.  This result is consistent
with that in JB04 for the 2dFGRS, and with our analysis of the 2PCF
in Paper I. Such a dependence on luminosity was
also found for the 2PCF by Norberg \etal (2002) using the 2dFGRS and
by Zehavi \etal (2005) using the SDSS.  The real-space power spectrum
$P(k)$ is approximately a power law for the range of $k$ considered
here.  The dotted lines in Figure \ref{fig:powreal} are the best
power-law fits to the measured $P(k)$ with the best-fitting parameters
indicated in each panel.  $P(k)$ decreases with $k$ approximately as
$k^{-1.4}$, with the slopes of the brightest samples being somewhat
shallower ($\sim k^{-1.2}$). Moreover, Figure \ref{fig:powreal}
shows that there is  a slight change in the slope of $P(k)$ at $k\approx 1\mpci$
for faint galaxies and at $k\approx 0.5\mpci$ for bright ones. 
Such a change in slope is much less pronounced in JB04, but it is evident
here due to the improvement in the observational data. As pointed out
by JB04, such a change can be understood as the transition between the
scales where the pair counts are dominated by galaxy pairs in the same
halo to those where galaxy pairs are mostly in separate halos. This
result may indicate that fainter galaxies tend to reside in smaller
halos, while brighter galaxies are found in massive halos, 
but a more detailed analysis is needed to explain the subtle
changes of the power spectrum shape with luminosity.
It is interesting to note that such a transition was also
found by Zehavi \etal (2004) in the projected two-point correlation function of galaxies.

\begin{figure*}
\vspace{-4.5cm}
\centerline{\psfig{figure=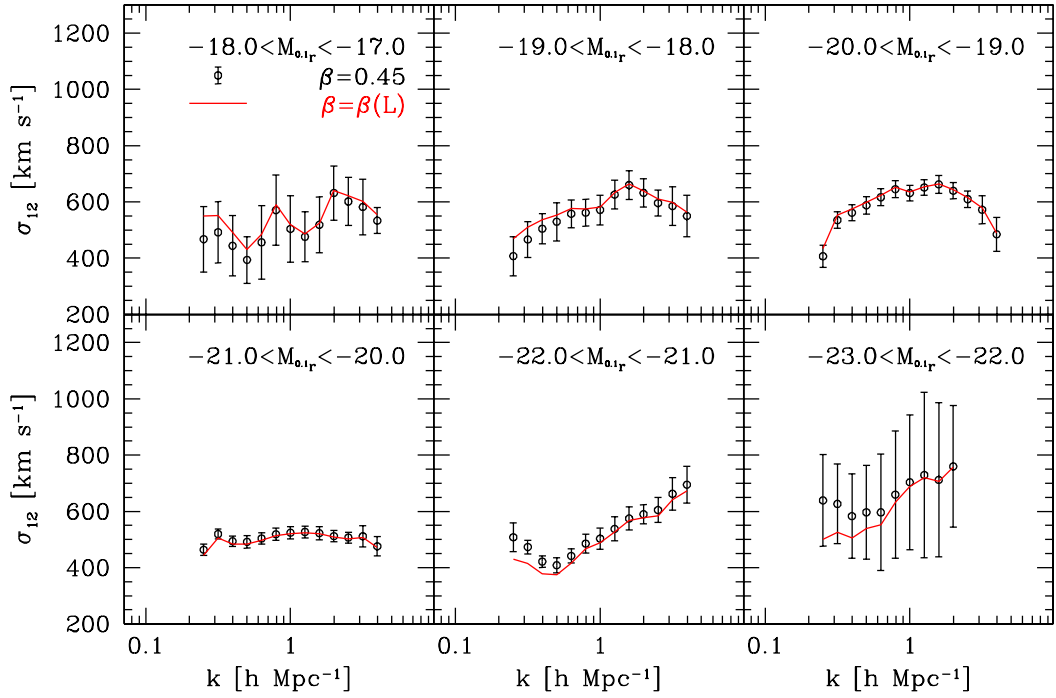,width=15cm}}
\vspace{-0.8cm}
\caption{The pairwise velocity dispersion $\sigma_{12}$
for galaxies in different luminosity intervals, as indicated.
The symbols with errorbars are for results that obtained by fixing
the redshift distortion parameter $\beta=0.45$, while
the lines are for those with $\beta$ varying with luminosity as
in Tegmark \etal (2004).
}
\label{fig:pvd_lum_beta}
\end{figure*}
\begin{figure}
\centerline{\psfig{figure=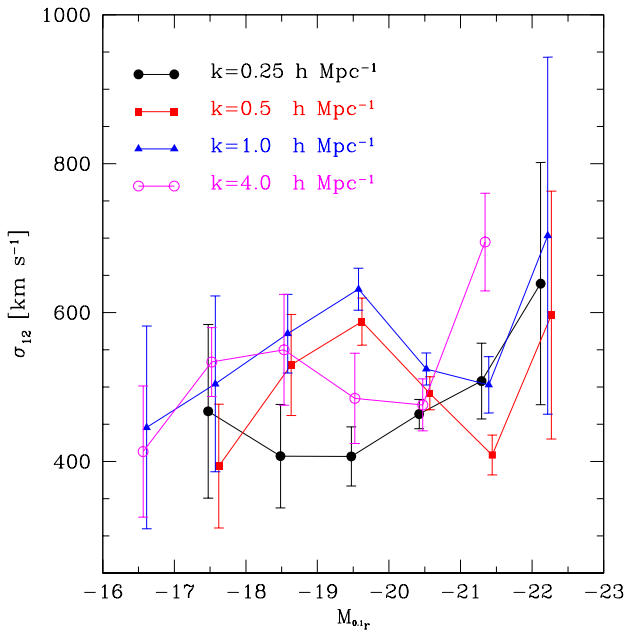,width=\hssize}}
\caption{The pairwise velocity dispersion $\sigma_{12}$ as a function
of luminosity, measured at $k = 0.25, 0.5, 1.0$ and $4.0 h$ Mpc$^{-1}$.
For clarity, small constants were added to the $M_{^{0.1}r}$ values for the curves
of $k = 0.25, 0.5$ and $4.0 h$ Mpc$^{-1}$.
}
\label{fig:pvdlum}
\end{figure}
\begin{figure}
\centerline{\psfig{figure=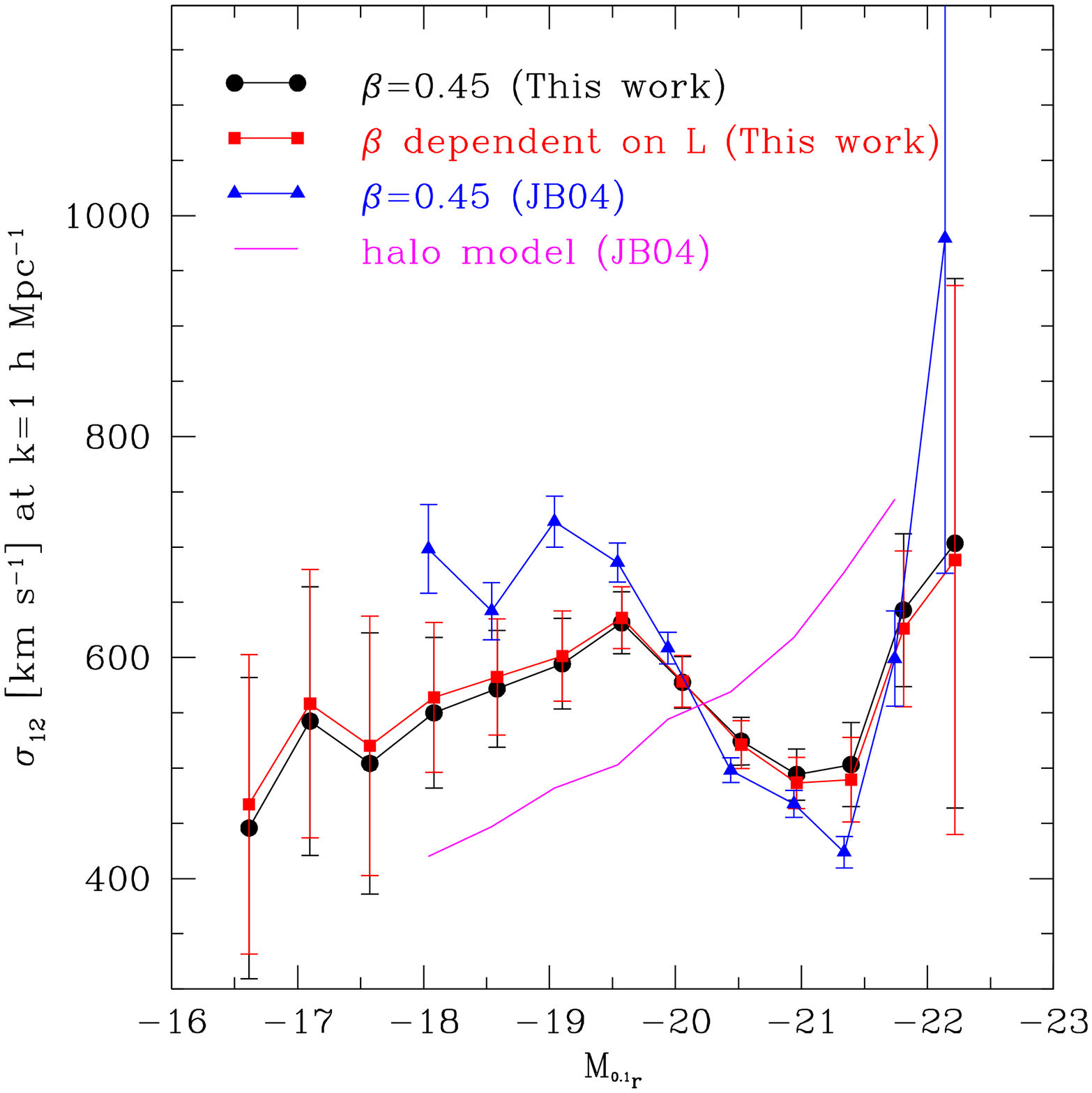,width=\hssize}}
\caption{The pairwise velocity dispersion $\sigma_{12}$ measured
at $k = 1 h$ Mpc$^{-1}$ in the SDSS ({\it circles}, $\beta=0.45$;
{\it squares}, $\beta$ varies with luminosity as in
Tegmark \etal [2004]),
compared with that in the 2dFGRS ({\it triangles}) presented by JB04.
The line without symbols shows the prediction of the HOD model of Yang \etal (2003).
}
\label{fig:2df}
\end{figure}

Fig.\ref{fig:pvd_lum_beta} shows the PVD $\sigma_{12}(k)$, which is
measured simultaneously with $P(k)$, for galaxies in different
luminosity intervals. {The errors of $\sigma_{12}(k)$ are also estimated
by bootstrap resampling technique.} It can be seen that for the $k$-values used
here, $\sigma_{12}(k)$ is a well-determined quantity.  As discussed in
\S1, there could be a luminosity dependence of the linear redshift
distortion parameter $\beta$, so a fixed value of $\beta$ might
have some impact on our results.  We investigate this by using the
luminosity dependence of the bias parameter
$b(M)/b^\ast=0.895+0.150(L/L^\ast)-0.040(M-M^\ast)$ given in Tegmark
\etal (2004) and $\beta^\ast=0.45$ (quantities with an
asterisk are defined to be  at the characteristic luminosity $L^\ast$).  The
results are plotted in Fig.\ref{fig:pvd_lum_beta} and
agree with those obtained for $\beta=0.45$ within the error bars. The largest
differences are seen  for bright galaxies on very large scales.  This indicates
that our $\sigma_{12}(k)$ measurements are robust to reasonable
changes of the $\beta$ values.

The dependence of the PVD on luminosity is clearly seen in this
figure. The observed PVD for $L^\ast$ galaxies (the
$-21<M_{^{0.1}r}<-20$ sample) is constant around 500$\kms$
at all scales. For faint galaxies, the PVD values increase as a
function of $k$, reach a maximum value of 650 km s$^{-1}$ at $k \simeq
1h$ Mpc$^{-1}$ and then decrease again.  For bright galaxies, the
behaviour is very different.  The PVD values decrease as a function of
$k$, reach a minimum value of 400 km s$^{-1}$ at $k \simeq 0.5 h$
Mpc$^{-1}$ and then increase.  It is interesting that these main
features agree with the findings by JB04 for the 2dFGRS, indicating
that the PVD as a function of the scale $k$ and the luminosity are
robustly determined with these surveys. As shown by JB04, these
features could not be reproduced by the halo model of Yang et
al. (2003), because the model predicts that the PVD increases
monotonically on small scales. Recently, Slosar, Seljak \& Tasitsiomi
(2005) made an attempt to interpret the observed luminosity dependence
of the PVD in the context of a halo model, and showed that the
luminosity dependence of the PVD can be predicted if some of the faint
galaxies are the satellite galaxies in high mass halos, which is
consistent with what JB04 expected as well as with galaxy formation
models and HOD modeling (e.g., Guzik \& Seljak 2002; Berlind \etal 2005;
Zehavi \etal 2005). However, a quantitative HOD model
that can simultaneously match the observations, such as the luminosity
functions, the correlation functions and the PVDs, still needs to be
worked out. The observed features of the PVD are very sensitive to
how the galaxies of different luminosity are distributed among dark
matter halos and also inside these halos. The PVD is therefore a good
constraint on both semi-analytical models of galaxy formation and the
halo occupation models for the galaxy distribution.

The luminosity dependence of the PVD is shown more clearly in Figure
\ref{fig:pvdlum}, where we have plotted $\sigma_{12}$ at $k=0.25,
0.5, 1$ and $4 \mpci $ for galaxies of different luminosities.  We again see
that the observed PVD for $M^\ast$ galaxies does not depend on  
$k$-value.  On large scales ($k=0.25\mpci$), $\sigma_{12}$ rises
as a function of increasing luminosity. 
However, on small scales, the PVD exhibits a 
well-defined minimum at luminosities  around $M^\ast-1$. 
The values of $\sigma_{12}$ at this minimum are  
$\simeq 500\kms$ at $k=1\mpci$ and $\simeq 400\kms$ at
$k=0.5\mpci$.  We thus conclude that  bright and faint galaxies  have
higher random motions than galaxies of intermediate luminosity.
The maximum relative velocities of faint galaxies ($\simeq 600 \kms$) occur on scales
$\sim 1\mpci$. For the brightest galaxies, the maximum
values occur on the smallest scales and reach  values close to  $1000\kms$. 

\begin{table}
\caption{The results of the PVD at $k=1\mpci$ of the luminosity subsamples}
\begin{center}
\begin{tabular}{cccccccc}
\hline\hline 
&Median Magnitude&$\sigma_v[\beta=0.45]$$^a$&$\sigma_v[\beta(L)]$$^b$\\
Sample&$M_{^{0.1}r}$&($\kms$)&($\kms$)\\
\hline 
L1...........& $-16.61$ & $446\pm 136$ & $467 \pm 135$\\ 
L2...........& $-17.10$ & $542\pm 122$ & $558 \pm 121$\\ 
L3...........& $-17.57$ & $504\pm 118$ & $520 \pm 117$\\ 
L4...........& $-18.08$ & $550\pm  68$ & $564 \pm  68$\\ 
L5...........& $-18.58$ & $572\pm  53$ & $582 \pm  53$\\ 
L6...........& $-19.10$ & $594\pm  41$ & $601 \pm  41$\\ 
L7...........& $-19.57$ & $631\pm  28$ & $636 \pm  28$\\ 
L8...........& $-20.06$ & $578\pm  23$ & $578 \pm  23$\\ 
L9...........& $-20.52$ & $524\pm  22$ & $521 \pm  22$\\ 
L10..........& $-20.96$ & $494\pm  23$ & $487 \pm  23$\\ 
L11..........& $-21.39$ & $503\pm  38$ & $489 \pm  38$\\ 
L12..........& $-21.81$ & $643\pm  69$ & $626 \pm  70$\\ 
L13..........& $-22.22$ & $703\pm 240$ & $688 \pm 248$\\ 
\hline
\multicolumn{8}{l}{$^a$ The PVD determined with $\beta=0.45$.}\\
\multicolumn{8}{l}{$^b$ The PVD determined with the luminosity dependence of $\beta$}\\
\multicolumn{8}{l}{\ \ \ taken into account.}
\end{tabular}
\end{center}
\end{table}
\begin{figure*}
\vspace{-5.5cm}
\centerline{\psfig{figure=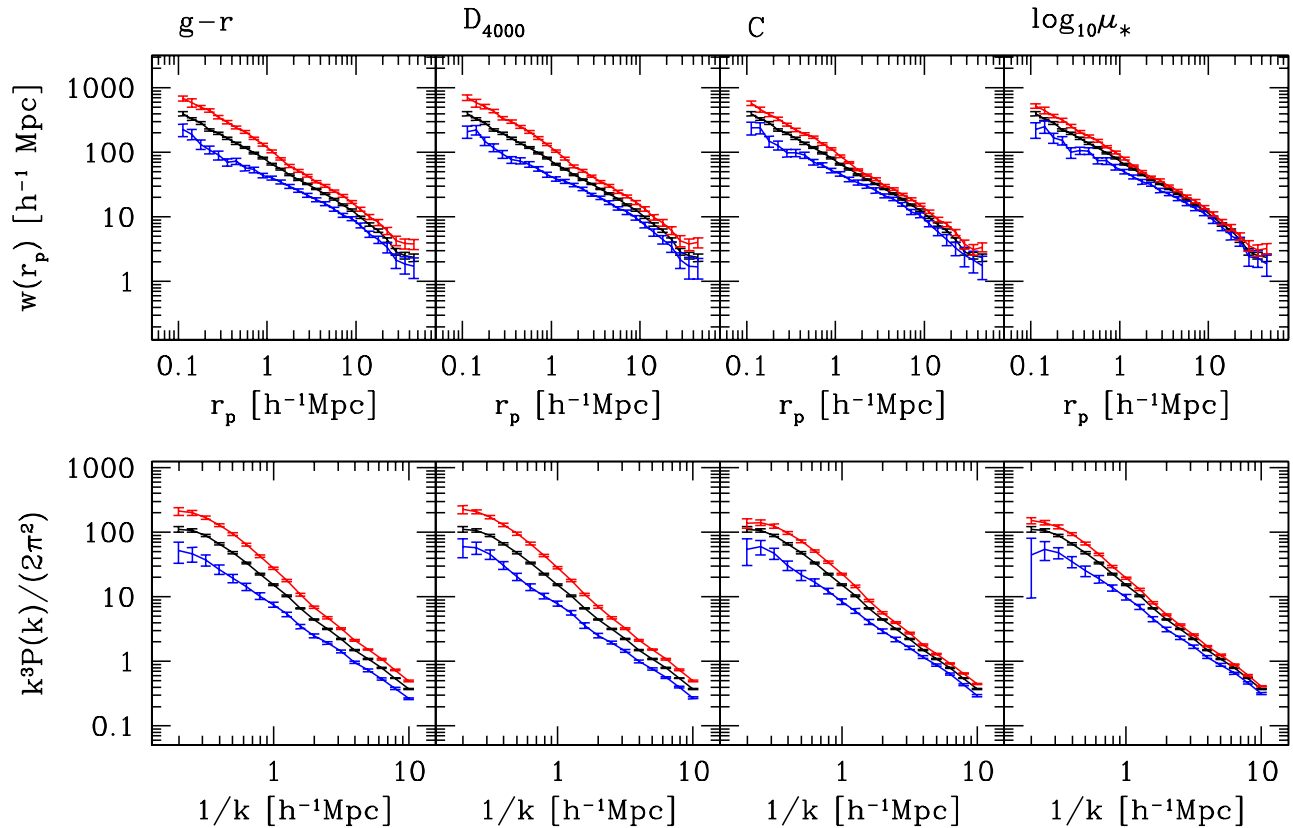,width=\hdsize}}
\vspace{-0.8cm}
\caption{The real space power spectrum $P(k)$ (bottom panels; here the
power spectra are given in the power-per-log-interval presentation:
$\Delta^2(k)=k^3P(k)/(2\pi^2)$) for galaxies in the luminosity interval of 
$-21<M_{^{0.1}r}<-20$ 
and with different properties (from left to right: $g-r$ colour, D$_{4000}$,
concentration and log stellar surface mass density $\log_{10}\mu_\ast$),
compared to the corresponding $w_p(r_p)$ measurements (top panels) from Paper I
(see the third row in Fig.10 of Paper I).
In each panel, the {\it black} is for the full sample, while the {\it red}
({\it blue}) is for the subsample with larger (smaller) value of the 
corresponding physical parameter.
}
\label{fig:pk}
\end{figure*}
\begin{figure*}
\vspace{-0.2cm}
\centerline{\psfig{figure=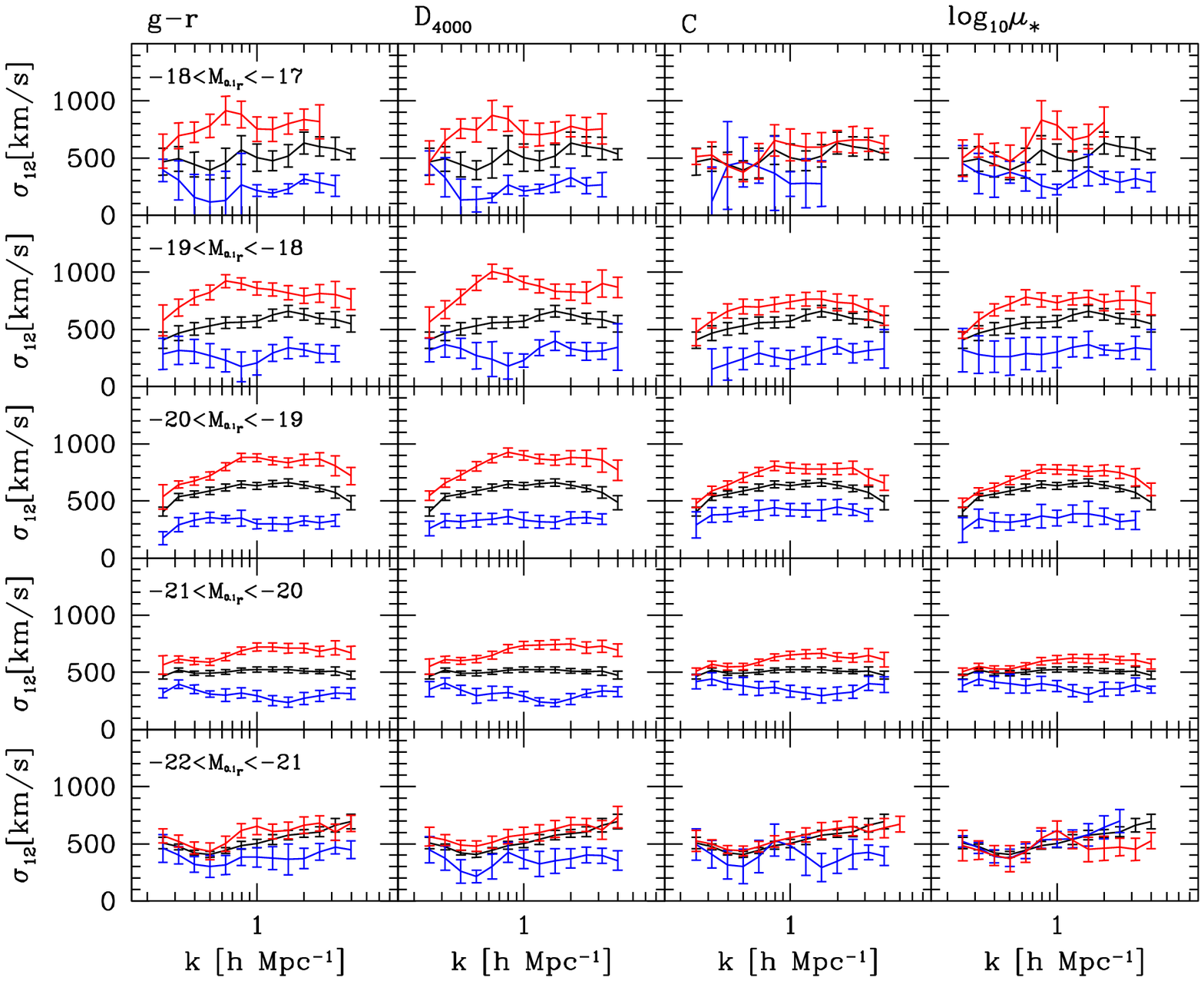,width=\hdsize}}
\vspace{-3.5cm}
\caption{The pairwise velocity dispersion $\sigma_{12}$ for galaxies
with various luminosities and physical properties.
Panels from top to bottom correspond to galaxies in different luminosity
intervals (as indicated in the first column), while panels from
left to right correspond to galaxies divided by different physical
quantities (as indicated above each column). In each panel,
{\it red} ({\it blue}) is for the subsample with larger (smaller) value
of the corresponding physical quantity, while {\it black} is for the
full sample.
}
\label{fig:pvd_lum}
\end{figure*}

It is interesting to compare the results presented in this paper with
those of JB04.  To do this, one must transform from the luminosities
measured in the $b_J$ band in the 2dFGRS to 
those measured in the  $^{0.1}r$ band in the SDSS. 
According to the luminosity functions in the two surveys
(Madgwick \etal 2002, Blanton \etal 2003), the average difference in
absolute magnitude in the two bands is $\sim 0^{m}.9$.  
In Figure \ref{fig:2df} we compare our SDSS results with those of the 2dFGRS
after taking into account this difference. 
It can be seen that the two measurements are
quite consistent with each other. It is worth noting that the
agreement is almost perfect for luminosities brighter than $-19.5$,
but for fainter luminosities, $\sigma_{12}$ is slightly smaller as
measured from the SDSS compared to 2dFGRS.
The latter may be due to the fact the volumes covered
by the faint samples are not sufficiently large (Mo, Jing \& B\"orner 1997).
The results obtained with varying
$\beta$ are also plotted in this figure.
We also plot the
prediction for the 2dFGRS PVD of JB04 based on the halo model of Yang
\etal (2003), which clearly does not match the observations.
The $\sigma_{12}$ measurements as a function of luminosity are listed in
Table 1.

\begin{figure*}
\vspace{-0.5cm}
\centerline{\psfig{figure=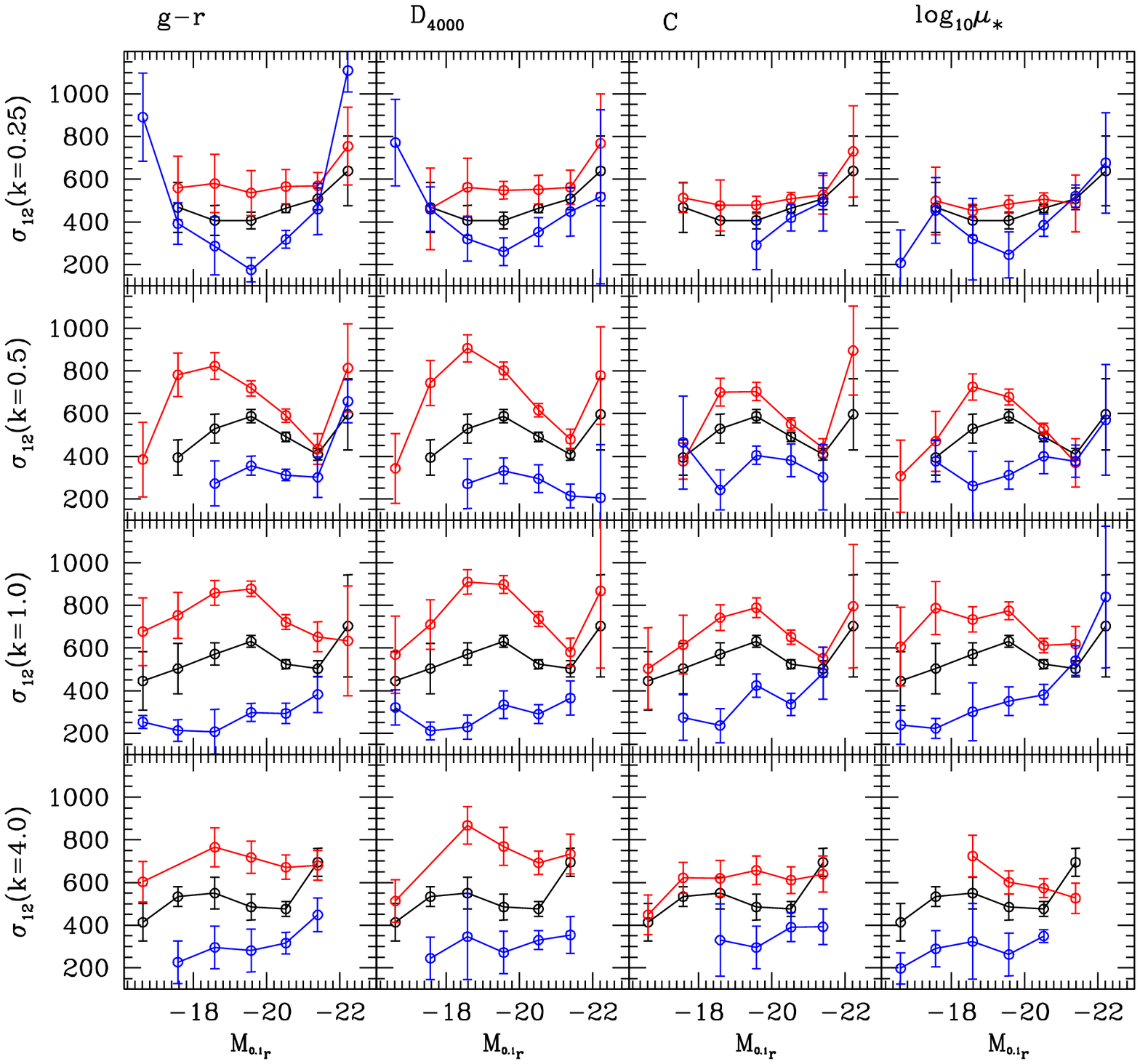,width=\hdsize}}
\vspace{-0.8cm}
\caption{The pairwise velocity dispersion $\sigma_{12}(k)$
as a function of absolute magnitude,
measured at $k=$0.25, 0.5, 1 and 4 $\mpci$
for different classes of galaxies.
Panels from top to bottom correspond to
measurements on different wavenumbers, while panels from left to right
correspond to galaxies divided by different physical quantities, as incidated
above each column. In each panel, {\it red} ({\it blue}) is for the
subsample with larger (smaller) value of the corresponding physical quantity,
while {\it black} is for the full sample.}
\label{fig:pvd_lum_par}
\end{figure*}
\subsection{The dependence on $g-r$, $D_{4000}$, $C$ and $\mu_\ast$}
To probe the dependence of the PVD on parameters related to the recent star formation
history of galaxies ($g-r$, D$_{4000}$) and on parameters related to
galaxy structure ($C$, $\mu_\ast$), we divide the galaxies into two
subclasses: galaxies with larger values of the given  physical quantity
(hereafter denoted as "red" galaxies) and galaxies with smaller values
(hereafter denoted as "blue" galaxies). These subsamples were
described in \S2 of Paper I. 

Fig.\ref{fig:pk} shows the real space power spectrum
$P(k)$ in the space of luminosity {\it vs} the physical quantities. The results are
compared with the measurements of 
$w_p(r_p)$ presented in Paper I. 
As can be seen, the results of $P(k)$ and $w_p(r_p)$ are qualitatively very similar.
Galaxies with redder colours,
stronger 4000\AA\ breaks, more concentrated structure and higher
surface densities have higher clustering amplitude    on all scales
and at all luminosities, with the difference more pronounced
on small scales and for faint galaxies.
Furthermore, the dependence on  parameters associated
with recent star formation is much stronger than the dependence on structural
parameters.  

\begin{figure*}
\vspace{-2.8cm}
\centerline{\psfig{figure=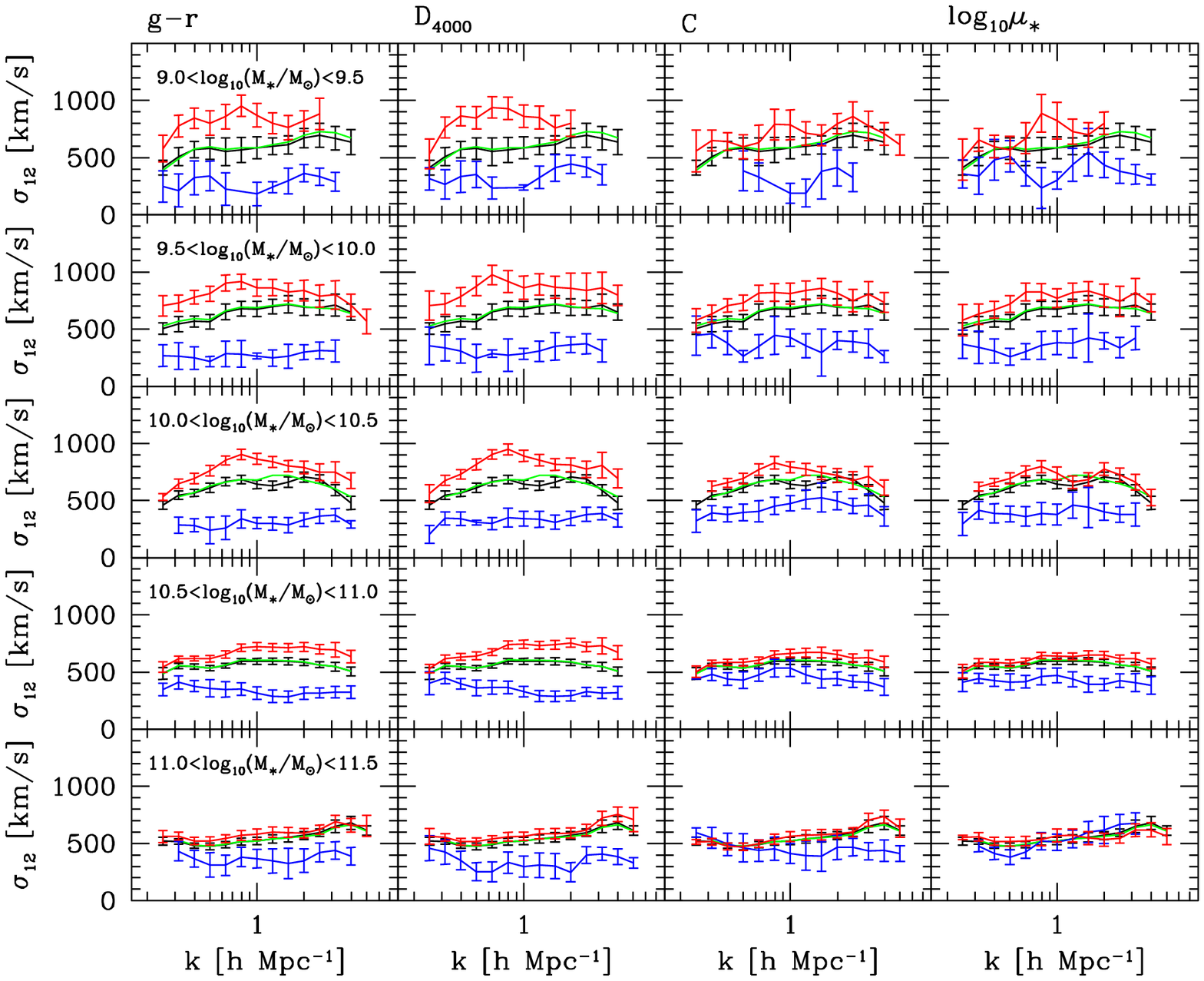,width=\hdsize}}
\vspace{-0.8cm}
\caption{The pairwise velocity dispersion $\sigma_{12}$ for galaxies
with various stellar masses and physical properties. 
The symbols are the same as in Fig.\ref{fig:pvd_lum}, except that 
the {\it green} line in each panel is for the full sample volume-limited
in stellar mass.}
\label{fig:pvd_smass}
\end{figure*}

\begin{figure*}
\vspace{-0.5cm}
\centerline{\psfig{figure=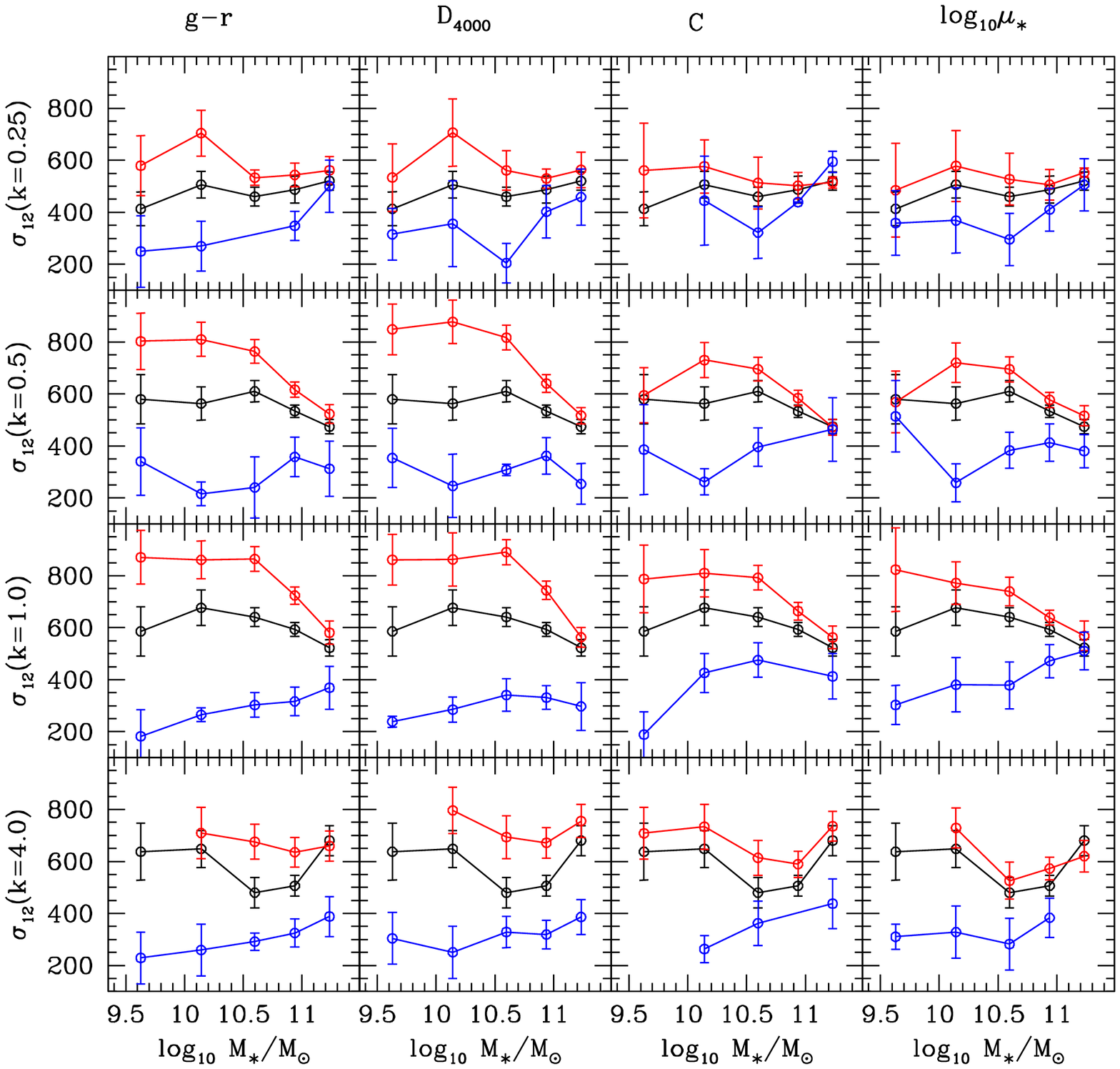,width=\hdsize}}
\vspace{-0.8cm}
\caption{The pairwise velocity dispersion $\sigma_{12}(k)$
as a function of stellar mass, measured at $k=$0.25, 0.5, 1 and 4 $h$ Mpc$^{-1}$
for different classes of galaxies. The symbles are the same as 
in Fig.\ref{fig:pvd_lum_par}.}
\label{fig:pvd_smass_par}
\end{figure*}

Fig.\ref{fig:pvd_lum} shows the dependence of the PVD $\sigma_{12}$
on $k$  for galaxies with  different luminosities and  physical properties.
The samples and the
symbols are the same as in Fig.\ref{fig:pk}.  In
Fig.\ref{fig:pvd_lum_par}, we plot $\sigma_{12}$ measured at $k=$0.25, 0.5,
1 and 4 $\mpci$ as a function of luminosity for galaxies
with different physical properties. 
Plots corresponding to Figs.\ref{fig:pvd_lum} and \ref{fig:pvd_lum_par},
as a function of stellar mass rather than luminosity are shown in
Figs.\ref{fig:pvd_smass} and \ref{fig:pvd_smass_par}.
We find that  red galaxy
populations as defined by both $g-r$ and D$_{4000}$ and early-type galaxy
populations as defined by $C$ and $\mu_\ast$ have larger relative
velocities on all scales and at all luminosities/masses than blue and
late-type populations.  We can also see that the dependence of
$\sigma_{12}$ on these physical properties is stronger for faint/low mass
galaxies and on small scales.  On very large scales,
galaxies of all  luminosities/masses have very similar  pairwise
peculiar velocities. 

%

\begin{figure}
\vspace{-0.5cm}
\centerline{\psfig{figure=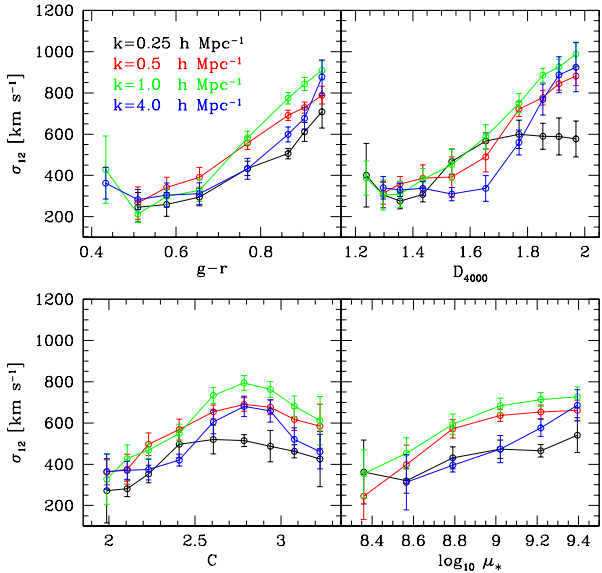,width=9cm}}
\vspace{-0.8cm}
\caption{The pairwise velocity dispersion $\sigma_{12}$ as
a function of various physical quantities, 
measured at $k=$0.25, 0.5, 1 and 4 $\mpci$ (as indicated
in the top-left panel).
All the samples are selected to lie in the stellar mass range $10<\log_{10}M_\ast<11$.}
\label{fig:pvd_par}
\end{figure}

In order to study the detailed dependence of the PVD on galaxy  properties,
we restrict our analysis to galaxies with stellar masses
in the range $10<\log_{10} M_\star < 11$, and
divide the sample into subsamples with finer bins in $g-r$, D$_{4000}$, 
$C$ and $\mu_*$ (see Table 3 of Paper I).
In Fig.\ref{fig:pvd_par} we show the pairwise velocity dispersion
$\sigma_{12}$ on different scales, as a function of these quantities.
The dependences of $\sigma_{12}$ on quantities related to past star formation
history and on quantities related to galaxy structure are quite different.
Fig.\ref{fig:pvd_par} shows that                                          
$\sigma_{12}$ increases with increasing $g-r$ or D$_{4000}$.
On the other hand,  $\sigma_{12}$ as a function of concentration  exhibits
a maximum at $C\sim 2.6$ for $k=0.25\mpci$ and at $C\sim 2.8$ for larger $k$ values,
reaching $\sim 700\kms$ on small
scales and $\sim 500\kms$ on large scales.
Furthermore, on very small scales, "star-forming" galaxies
with $D_{4000}<1.5$ have nearly constant peculiar velocities
($\sim  300\kms$), while $\sigma_{12}$ increases sharply above
$D_{4000}=1.5$, reaching $\sim 1000\kms$ for galaxies with $D_{4000}=2$.
The PVD is an indicator of the depth of the local gravitational
potential. Therefore, we are led to the 
conclusion that the reddest galaxies
move in the strongest gravitational fields. 
A substantial fraction of these red galaxies  must reside in clusters, 
while most galaxies with blue colours, recent star formation, 
and diffuse structure populate the field.
Interestingly, the behaviour of $\sigma_{12}$ as a function of concentration
is more complicated. Galaxies with {\em intermediate} value of $C$
have the highest relative velocities. This may indicate that many
red galaxies in clusters are really spirals whose ongoing star formation
has been truncated.

\section{Discussion and Conclusions}
In this paper we have made a detailed study of the power spectrum and
the pairwise velocity dispersions of galaxies classified according to
their luminosity, stellar mass, colour, 4000\AA\ break, concentration
index, and  stellar surface mass density, using the SDSS DR2.
Our results can be summarized as follows:
\begin{enumerate}
\item The real space power spectrum and the pairwise velocity
dispersion as a function of luminosity in the SDSS are in good
agreement with the result presented by Jing \& B\"orner (2004) for the
2dFGRS, after taking into account the photometric bandpass  difference between
the two surveys.
\item Galaxies with redder colours, stronger 4000\AA\ breaks, more
concentrated structure and higher surface mass densities have 
larger clustering power and larger relative velocities on all scales
and at all luminosities/stellar masses.
\item The dependences of the clustering power and PVD on the
parameters related to recent star formation are stronger than those on
the parameters related to galaxy structure, especially on small
scales and for faint galaxies.
\item The reddest galaxies and galaxies of intermediate concentrations move
in the deepest gravitational field. A large fraction of these objects
must reside in clusters, while most galaxies with blue colours,
recent star formation, and diffuse structure
populate the field.
\end{enumerate}

From the above results, it is not difficult to understand why the
PVD on small scales ($k=1\mpci$) exhibits a minimum at $M^\ast-1$,
and why faint galaxies have larger relative velocities than bright  
galaxies (JB04). As seen both in Fig.\ref{fig:pk} in this paper and
in Fig.10 of Paper I, 
faint red galaxies are more strongly clustered than  faint blue galaxies,
and have comparable or even greater clustering power than  bright
red galaxies.  However, the majority of faint galaxies have
blue colours and cluster weakly. The fraction of
red galaxies is only about 30 percent for luminosities fainter than
$-19$ (Table 1 of Paper I). This means that  the power spectrum of 
faint galaxies is dominated by the blue population, which leads
to a monotonic increase of the clustering strength with luminosity
(Norberg \etal 2001; Zehavi \etal 2005; Tegmark \etal 2004).
However, the faint red population in rich clusters, even though small 
in number, can dominate the PVD on small scales,
because the PVD is more sensitive to the galaxy population in rich clusters
than the power spectrum (see Mo, Jing \& B\"orner [1997] for a discussion). This is
also clearly seen in Figure \ref{fig:pvd_lum_par}. Our results
therefore support the conjecture made by JB04 that a large fraction 
of the faint galaxy population must reside in rich clusters.
We demonstrate that this fraction is predominantly red.

Although most of the results presented in this paper  can be 
understood qualitatively by considering
known trends in galaxy properties as a function
of environment, a quantitative understanding requires galaxy formation models 
(e.g. Kauffmann \etal 1993, 1997, 1999; Cole \etal 1994,2000; Somerville \&
Primack 1999;  Kang \etal 2005;
Croton \etal 2006) or  halo occupation models (e.g. Jing, Mo, B\"orner 1998; Seljak 2000;
Berlind \& Weinberg 2002; Cooray \& Sheth 2002;
Yang \etal 2003; Zheng 2004; Slosar, Seljak, Tasitsiomi 2006).
Through such studies, the results in
this paper will provide stringent constraints on these models, and will
provide important clues where galaxies of different physical
properties are located and how they have formed and evolved.

\section*{Acknowledgments}

We are grateful to Dr. Michael Blanton for his help with the
NYU-VAGC. We thank the SDSS teams for making their data publicly
available and the referee for a detailed report. This work
is supported by NKBRSF(G19990754), by NSFC(Nos.10125314, 10373012,
10073009), by Shanghai Key Projects in Basic research (04jc14079,
05xd14019), by the Max Planck Society, and partly by the Excellent
Young Teachers Program of MOE, P.R.C.  CL acknowledges the financial
support of the exchange program between Chinese Academy of Sciences
and the Max Planck Society.

Funding for the creation and distribution of the SDSS Archive has been
provided by the Alfred P. Sloan Foundation, the Participating
Institutions, the National Aeronautics and Space Administration, the
National Science Foundation, the U.S. Department of Energy, the
Japanese Monbukagakusho, and the Max Planck Society. The SDSS Web site
is http://www.sdss.org/.
                                                                                
The SDSS is managed by the Astrophysical Research Consortium (ARC) for
the Participating Institutions. The Participating Institutions are The
University of Chicago, Fermilab, the Institute for Advanced Study, the
Japan Participation Group, The Johns Hopkins University, Los Alamos
National Laboratory, the Max-Planck-Institute for Astronomy (MPIA),
the Max-Planck-Institute for Astrophysics (MPA), New Mexico State
University, Princeton University, the United States Naval Observatory,
and the University of Washington.

%


\end{document}